\documentclass[12pt]{article}
\usepackage{amsmath,amssymb}
\usepackage{bm}
\usepackage{graphicx}
\usepackage{authblk}

\title{Discrete symmetries and the Lieb-Schultz-Mattis theorem}
\author[1,2]{Takaichi Isoyama}
\author[1]{Kiyohide Nomura}
\affil[1]{
Department of Physics, Kyushu University,
Fukuoka 819-0395 JAPAN}
\affil[2]{present address:
Nippon Light Metal Company,Ltd.,JAPAN}
\date{}
\begin{document}

\maketitle

\begin{abstract}
In this study, we consider one-dimensional (1D) quantum spin systems with
translation and discrete symmetries (spin reversal,
space inversion and time reversal symmetries).
 By combining the continuous U(1) symmetry with the discrete symmetries
 and using the extended
 Lieb-Schultz-Mattis theorem \cite{Lieb-Schultz-Mattis-1961}\cite{Nomura-Morishige-Isoyama-2015}, we investigate
 the relation between the ground states, energy spectra and
 symmetries.
 For half-integer spin cases, we generalize the dimer and N\'eel
concepts using the discrete symmetries,
 and we can reconcile the LSM theorem with the dimer or N\'eel states,
 since there was a subtle dilemma.
 Furthermore, a part of discrete symmetries is enough
 to classify possible phases. 
 Thus we can deepen our understanding of the relation between the LSM theorem and 
 discrete symmetries.
 \end{abstract}

\section{Introduction}

In many body quantum systems,
it is important to investigate energy spectra,
that is, whether gapless or gapped,
or the degeneracy of ground states. 
The structure of energy spectra is related with other physical
properties, such as the symmetry breaking.
Therefore,
in nonrelativistic many body quantum systems,
general and rigorous theorems remain important.
The Lieb-Schultz-Mattis (LSM) theorem \cite{Lieb-Schultz-Mattis-1961} is one of them. 

Historically,
Lieb, Schultz and Mattis treated
the $S=1/2$ XXZ spin chain,
and they showed there exists a low energy $O(1/L)$ ($L$: system size)
excited state above the ground state. 
Affleck and Lieb \cite{Affleck-Lieb-1986} studied
general spin $S$ and SU($N$) symmetric cases,
and showed the same result as the LSM theorem for half-integer $S$.
In addition, they considered the relation between the space inversion and
the spin reversal symmetries.
Independently, using the twisted boundary condition,
Kolb \cite{Kolb-1985} showed,
for half-integer spin chains, the nontrivial periodicity of the wave
number $q\rightarrow q+\pi$
of the lowest energy dispersion in the zero magnetization subspace.
And he discussed the continuity of the energy dispersion for $q$.
One limitation of the traditional LSM theorem
\cite{Lieb-Schultz-Mattis-1961,Affleck-Lieb-1986, Oshikawa-Yamanaka-Affleck-1997}
is the assumption of the unique ground state
(or the unique lowest energy state 
in the fixed magnetization subspace)
for the finite system,
which is violated in several cases with frustrations
(Majumdar-Ghosh model \cite{Majumdar-Ghosh-1969,Majumdar-Ghosh-1970} and the incommensurate region).
We have separated
the proof process of the LSM theorem
from the assumption of the unique ground state \cite{Nomura-Morishige-Isoyama-2015}.

In our previous paper, 
we proved that the lowest energy spectrum is a continuous function of
the wave number $q$
for the irrational magnetization,
and the nontrivial periodicity of $q$
for the rational magnetization
(the latter statement is a generalization of \cite{Oshikawa-Yamanaka-Affleck-1997}). 
Here we mean ``rational magnetization'',
to be the case where the magnetization per site is
a relatively simple rational number
in the thermodynamic limit:
\begin{align*}
\lim_{L\rightarrow \infty} \frac{S^{z}_{T}}{L}
 =\frac{m}{n}
\end{align*}
($L$: system size, $S^{z}_{T} \equiv \sum_{j} S^{z}_{j}$, $m$ and $n$ are
coprimes),
and ``irrational magnetization'' otherwise. 
Thus, there are two possibilities:
 in the first case, the energy spectrum is continuous with $q$, so it doesn't
 have an energy gap (see Fig.1 (a));
 secondly, an energy gap exists between the ground state and
 continuous excited states (see Fig 1 (b)).
 
\begin{figure}[h]
\begin{center}
\includegraphics[height=30mm]{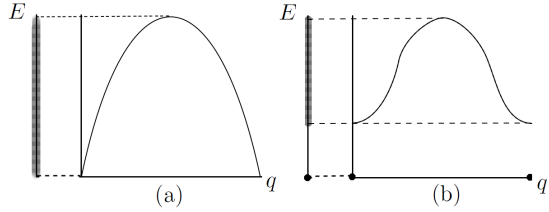}

\caption{Example of energy spectra:cases (a),(b).
Left side bar is side view of spectrum expressing energy distribution.}
\end{center}
\end{figure}

In a half-integer spin chain with U(1) symmetry at zero magnetization, 
when the energy spectrum is gapless (case (a)),
the low energy modes are at $q=0,\pi$.
When the energy spectrum is gapped (case (b)),
several possibilities can
be considered as the ground states.
One candidate is the dimer state: 
\begin{align*}
|\Psi_{\rm dimer}\rangle\equiv[1,2][3,4]\dots [L-1,L],
\quad
[i,i+1]=\frac{1}{\sqrt{2}}(|\uparrow\rangle_i|\downarrow\rangle_{i+1}-|\downarrow\rangle_i|\uparrow\rangle_{i+1}),
\end{align*}
and a counterpart produced from the translation by one site. 
Another one
is the N\'{e}el state: 
\begin{align*}
|\Psi_{\rm N\acute{e}el}\rangle\equiv|\uparrow_1\downarrow_2\uparrow_3\dots\downarrow_L\rangle
\end{align*}
and its counterpart.
From these states, we can construct the $q=0,\pi$ wave number states.
The $q=0$ states for both cases have a common feature
of the discrete symmetries
(spin reversal, space inversion and time reversal),
whereas at $q=\pi$ the N\'eel-like state has different discrete symmetries
from those of the dimer-like state.

On the other hand,
the LSM-type variational state at $q=\pi$ contains
{\em both N\'eel-like} (see Eq. (\ref{eq:n1})) {\em and the dimer-like}
(see Eq. (\ref{eq:d1})){\em components}. 
This suggests
that both states should be always 
degenerate with the $q=0$ ground state, which seems unnatural.
In fact, in the Majumdar-Ghosh model,
only the dimer-like state at $q=\pi$ is degenerate with the $q=0$ ground state
\cite{Majumdar-Ghosh-1969,Majumdar-Ghosh-1970}.
In this study,
considering the discrete symmetries,
we will show that the normalization is important to resolve this dilemma. 
Furthermore, we show that a part of the discrete symmetries is
enough to distinguish the N\'eel-like and dimer-like
states, and for the normalization discussion.
One of the interesting applications is the magnetic plateau 
\cite{Oshikawa-Yamanaka-Affleck-1997},
to classify the possible phases.

The layout of this paper is as follows.
In Sect. \ref{section:symmetries}, we define the symmetries 
(rotation, space inversion, translation
and time reversal) and relating operators,
then we discuss the symmetry properties of the N\'eel or dimer states.
Section \ref{section:Lieb-Schultz-Mattis} is a review of the LSM theorem and twisting operators.
In Sect. 
\ref{section:Discrete symmeties and LSM theorem}, we construct
several states with wave number $q=\pi$ 
by acting the twisting operator to the $q=0$ ground state,
then we study the discrete symmetries of the constructed $q=\pi$ states.
When U(1) symmetry holds, these $q=\pi$ states have the $O(1/L)$ low energy from
the LSM theorem.
Section \ref{section:normalization} is our main result;
we will explain why some of the $q=\pi$
states do not have to satisfy the LSM inequality,
considering the normalization of the states.
In Sect. \ref{section:partial-discrete-symmetry},
we show that a part of discrete symmetries
is enough for the above arguments.
And we apply our result to several examples ($S=1/2$ NNN XXZ chain,
spin ladder, spin tube and distorted diamond chain) in Sect. \ref{section:examples}.
In Sect. \ref{section:ground-state-symmetries}, we discuss symmetric properties of the ground state
with system sizes $L=4n,4n+2,2n+1$.
Section \ref{section:summary} is a summary and discussion.

 \section{Symmetries and operators
\label{section:symmetries}
 }

In this section, we explain the symmetries and the operators
of the many body system.
As an example, we treat a 1D spin system:
\begin{align}
\hat{H} = \sum_{j=1}^L\sum_{r=1}^{L/2}J_x(r)\hat{S}^x_j\hat{S}^x_{j+r}+J_y(r)\hat{S}^y_j\hat{S}^y_{j+r}+J_z(r)\hat{S}^z_j\hat{S}^z_{j+r},
\label{eq:xyz}
\end{align}
where $(\hat{\bm{S}}_j)^2=S(S+1)$. 
In addition, the system has a periodic boundary
 condition
($\hat{S}^{x,y,z}_{L+j}=  \hat{S}^{x,y,z}_{j}$)
 and the system size $L$ is $4n$($n$:integer).

 \subsection{Rotational operator}
 We define rotational operators around the $x, y, z$ axes:
\begin{align}
\hat{U}^x_{\theta}\equiv\exp{(-i\theta \hat{S}^x_T)},~\hat{S}^x_T\equiv\sum^L_{j=1}\hat{S}^x_j,\\
\hat{U}^y_{\theta}\equiv\exp{(-i\theta \hat{S}^y_T)},~\hat{S}^y_T\equiv\sum^L_{j=1}\hat{S}^y_j,\\
\hat{U}^z_{\theta}\equiv\exp{(-i\theta \hat{S}^z_T)},~\hat{S}^z_T\equiv\sum^L_{j=1}\hat{S}^z_j.
\end{align}
We can calculate, e.g.,
\begin{align}
 (\hat{U}^x_{\theta})^{\dagger}\hat{S}^y_k\hat{U}^x_{\theta}
 &= \hat{S}^y_{k} + i\theta[\hat{S}_T^x,\hat{S}^y_{k}] 
	+\frac{1}{2!}\left(i\theta\right)^2[\hat{S}_T^x,[\hat{S}_T^x,\hat{S}^y_{k}]]
 +\cdots
\notag\\
 &= \cos{(\theta)}\hat{S}^y_k-\sin{(\theta)}\hat{S}^z_k.
 \label{spinrotation}
\end{align}

\subsection{Spin reversal operators}
In spin rotational operators, the $\theta=\pi$ case (spin reversal operator)
is important.
Since $(\hat{U}^z_{\pi})^2=\exp(-2\pi i\hat{S}_T^z)=1$, the eigenvalue
of $\hat{U}^z_{\pi}$ is $\pm 1$ (because
the eigenvalue of $\hat{S}_T^z$ is integer for even $L$).  
In the same way, $(\hat{U}^x_{\pi})^2 = (\hat{U}^y_{\pi})^2 =1$.

\subsection{Space inversion operators}
  
 We define two kinds of space inversion operators;
site inversion (inversion on the lattice site),
and link inversion (inversion on the link):
\begin{align}
\mbox{\ (site inversion)\ }:\hat{P}^{\dagger}_{\rm
 site}\hat{S}^{x,y,z}_{j}\hat{P}_{\rm site} &=\hat{S}^{x,y,z}_{L-j},
 \\
\mbox{\ (link inversion)\ }:\hat{P}^{\dagger}_{\rm
 link}\hat{S}^{x,y,z}_{j}\hat{P}_{\rm link} &=\hat{S}^{x,y,z}_{L-j+1},
\end{align}
where
$\hat{P}^{\dagger}_{\rm
 site}\hat{S}^{x,y,z}_{L}\hat{P}_{\rm site} =\hat{S}^{x,y,z}_{L}$
 and
 $\hat{P}^{\dagger}_{\rm
 site}\hat{S}^{x,y,z}_{L/2}\hat{P}_{\rm site} =\hat{S}^{x,y,z}_{L/2}$.

Since $\hat{P}_{\rm site}^2=\hat{P}_{\rm link}^2=\hat{1}$,
the eigenvalues of space inversions are $\pm 1$.

 \subsection{Translation operator}
 We define the translation operator by one site as
 \begin{align}
\hat{U}^{\dagger}_{\rm trl}\hat{S}^{x,y,z}_{j}\hat{U}_{\rm trl}=\hat{S}^{x,y,z}_{j+1},
\end{align} 
and the wave number $q$
as the eigenvalue of the translation operator:
\begin{equation}
\hat{U}_{\rm trl}|S^z_T;q\rangle=e^{iq}|S^z_T;q\rangle.
\end{equation}
The wave number $q$ is periodic as $q\rightarrow q+2\pi$
(Brillouin zone).

\subsection{Time reversal operator}

We introduce the time reversal operator $\hat{K}$,
which has the following properties:
for spin operators
\begin{align}
\hat{K}^{\dagger}\hat{\bm{S}}\hat{K}=-\hat{\bm{S}},
\label{eq:Time-reversal-operator-1}
\end{align}
and it is an antilinear operator, i.e., for a complex number $c$
\begin{align}
 \hat{K}^{\dagger} c \hat{K}= c^{*}.
 \label{eq:Time-reversal-operator-2}
\end{align}
The operator $\hat{K}$ commutes with the rotation, space
inversion, and translation operators; however, it does not commute with the twisting
operator (see Appendix \ref{appendix:Time-reversal-operator}).

  \subsection{Ground state properties
\label{section:ground-state}
  }
It is natural to assume that
at least one of the ground states $|\Psi_g\rangle$ has the following symmetry properties:
\begin{align}
\hat{U}_{\rm trl}|\Psi_g\rangle &=|\Psi_g\rangle, \;
\hat{P}_{\rm site} | \Psi_{g} \rangle = \hat{P}_{\rm link} | \Psi_{g} \rangle =
 | \Psi_{g} \rangle,
\notag \\
\hat{U}^{y}_{\pi} | \Psi_{g} \rangle &= \hat{U}^{z}_{\pi} | \Psi_{g}
 \rangle = | \Psi_{g} \rangle, \;
\hat{K}  | \Psi_{g} \rangle =  | \Psi_{g} \rangle
 \label{eq:ground-state-condition}
\end{align}
(furthermore in the case of U(1) symmetry around the $z$-axis,
we assume $\hat{S}^{z}_{T} |\Psi_{g} \rangle =0$).
More precisely on the above statement,
we consider that
there are a finite number of energy eigenstates
in a range of $O(1/L)$ above the lowest energy,
and the state with symmetry eigenvalues (\ref{eq:ground-state-condition}) is contained in them.

For nonfrustrated cases,
these symmetries
(\ref{eq:ground-state-condition})
for the ground state can be proven
by the Marshall-Lieb-Mattis (MLM) theorem with system size $L=4n$ (see Sect. \ref{section:ground-state-symmetries}).
Even in the frustrated case 
the above statement may remain valid,
e.g.,
one of the ground states of the Majumdar-Ghosh model
satisfies (\ref{eq:ground-state-condition}).

The state $|\Psi_g\rangle$ is normalized as $\langle\Psi_g|\Psi_g\rangle=1$.

\subsection{Symmetries and eigenvalues of dimer and N\'eel states}

In this subsection, we consider the symmetry properties of the N\'eel state
or the dimer state.
First, we check the symmetry eigenvalues of $q=0$ states
\begin{align}
 |\Psi_{\rm N\acute{e}el1}\rangle+|\Psi_{\rm N\acute{e}el2}\rangle 
\label{eq:q01}
 \\
|\Psi_{\rm dimer1}\rangle+ |\Psi_{\rm dimer2}\rangle 
\label{eq:q02}
\end{align}
where 
\begin{align}
 |\Psi_{\rm N\acute{e}el1}\rangle\equiv|\uparrow_1\downarrow_2\uparrow_3\dots\downarrow_L\rangle,
  \quad
|\Psi_{\rm N\acute{e}el2}\rangle\equiv|\downarrow_1\uparrow_2\downarrow_3\dots\uparrow_L\rangle,
\label{neelstate}
\end{align}
and
\begin{align}
 & |\Psi_{\rm dimer1}\rangle\equiv[1,2][3,4]\dots [L-1,L],
 \quad
  |\Psi_{\rm dimer2}\rangle\equiv[2,3][4,5]\dots [L,1],
  \notag \\
&[i,i+1]=\frac{1}{\sqrt{2}}(\uparrow_i\downarrow_{i+1}-\downarrow_i\uparrow_{i+1}).
\label{neelanddimer2}
\end{align}
Both Eq. (\ref{eq:q01}) and Eq. (\ref{eq:q02}) have
$\hat{P}_{\rm site}=1,\hat{P}_{\rm link}=1,\hat{U}^y_{\pi}=1,\hat{K}=1$.

On the contrary,
the $q=\pi$ states,
constructed from the dimer or the N\'{e}el states, 
have different symmetry eigenvalues, so we can distinguish these two states.
The N\'{e}el-like state
\begin{align}
|\Psi_{\rm N\acute{e}el-like}\rangle\equiv |\Psi_{\rm
 N\acute{e}el1}\rangle-|\Psi_{\rm N\acute{e}el2}\rangle
 \label{neellikeqpi}
\end{align}
has eigenvalues
$q =\pi,\hat{P}_{\rm site}=1,\hat{P}_{\rm link}=-1,\hat{U}^y_{\pi}=-1,\hat{K}=-1$.
In contrast, dimer-like state
\begin{align}
|\Psi_{\rm dimer-like}\rangle\equiv |\Psi_{\rm dimer1}\rangle
 -|\Psi_{\rm dimer2}\rangle
 \label{dimerlikeqpi}
\end{align}
has eigenvalues
$q = \pi,\hat{P}_{\rm site}=-1,\hat{P}_{\rm link}=1,\hat{U}^y_{\pi}=1,\hat{K}=1$.

In this subsection we have discussed the symmetries of particular cases 
Eq. (\ref{neelstate}) and Eq. (\ref{neelanddimer2}).
In the following,
we will generalize 
these symmetry consideration, for wide models,
Eq. (\ref{eq:xyz}), and quasi 1D spin systems, etc.

 \section{The LSM theorem
\label{section:Lieb-Schultz-Mattis}
 }

In this section we review the LSM theorem.
The assumptions of the LSM theorem are U(1) symmetry,
translational symmetry and short-range interaction. 
For the 
Hamiltonian (\ref{eq:xyz}) it means $J_x(r)=J_y(r) \propto \exp{(-r/\xi)}$.

In \cite{Nomura-Morishige-Isoyama-2015}
we have proven the following inequality:
\begin{align}
& E(S^z_T,q+\frac{2\pi}{L}S^z_T+2\pi S)+E(S^z_T,q-\frac{2\pi}{L}S^z_T+2\pi
 S)-2E(S^z_T,q) 
 \notag\\
& \leq
\langle S_T^z;q|(\hat{U}_{2\pi}^{tw})^{\dagger}\hat{H}\hat{U}_{2\pi}^{tw}+(\hat{U}_{-2\pi}^{tw})^{\dagger}\hat{H}\hat{U}_{-2\pi}^{tw}-2\hat{H}|S_T^z;q\rangle \leq O(1/L),
\label{eq:elsmp}
\end{align}
where $\hat{U}^{\rm tw}_{2\pi l}$
($l$: integer)
are the twisting operators given by
\begin{equation}
 \hat{U}^{\rm tw}_{2\pi l}
  \equiv\exp\left(-\frac{2\pi
  i l}{L}\sum^L_{j=1}j(\hat{S}^z_j-S)\right),
%  \quad (l:\text{integer})
 \label{eq:twisting-operator-site}
\end{equation}
and $E(S_T^z;q)$ is one of the lowest energies in the $S_T^z; q$ subspace.
We introduced
the phase factor
$\exp(\frac{2\pi i l}{L} S \sum_{j=1}^{L}j)$
in Eq. (\ref{eq:twisting-operator-site}),
which was absent in the original twisting operator in\cite{Lieb-Schultz-Mattis-1961,Affleck-Lieb-1986},
in order to avoid cumbersome treatments of the edge term
$\exp(2 \pi i l \hat{S}^{z}_{1})$,
appearing from the translation operation on the original twisting operator
(see also Eq. (\ref{eq:TBC-translation-2pi-3}) in Appendix \ref{appendix:Translation-twisting-operator}).

One result of the extended LSM
theorem\cite{Nomura-Morishige-Isoyama-2015} is that if the magnetization
is irrational, then the energy spectra must be continuous for $q$.
Secondly,
by using Eq. (\ref{eq:elsmp}) and the squeezing technique, 
we get the next inequality for spin $S$ half-integer:
\begin{align}
|E(S_T^z=0;q) - E(S_T^z=0;q+\pi)|\leq O(1/L).
\label{eq:elsmine}
\end{align}

\subsection{Space inversion, spin reversal, and the twisting operator}

For the site space inversion, spin reversal and the twisting operator,
there are the relations
\cite{Nomura-Morishige-Isoyama-2015,Affleck-Lieb-1986}
\begin{align}
\label{pandtw1}\hat{P}_{\rm site}\hat{U}^{\rm tw}_{2\pi l}=\hat{U}^{\rm tw}_{-2\pi l}\hat{P}_{\rm site},\\
\label{uytw}\hat{U}^y_{\pi}\hat{U}^{\rm tw}_{2\pi
 l}=(-1)^{2S l}\hat{U}^{\rm tw}_{-2\pi l}\hat{U}^y_{\pi}.
\end{align}

However,
in the link space inversion operation
on the twisting operator $\hat{U}^{\rm tw}_{2\pi l}$,
 \begin{align}
  \hat{P}_{\rm link} \hat{U}^{\rm tw}_{2\pi l} \hat{P}_{\rm link}
  =\exp\left(-\frac{2\pi i l}{L} \left(\hat{S}^{z}_{T}- S L\right)\right) \hat{U}^{\rm tw}_{-2\pi l},   
 \end{align}
 the prefactor may be confusing with the translation operation on $\hat{U}^{\rm tw}_{2\pi l}$
 (see Appendix \ref{appendix:Translation-twisting-operator}).   
Therefore, we introduce another twisting operator
$\hat{U}^{\rm twl}_{2\pi l}$ 
as \cite{Nomura-Kitazawa-1998}
\begin{align}
 \hat{U}^{\rm twl}_{2\pi l}
 \equiv\exp\left(-\frac{2\pi i
 l}{L}\sum^L_{j=1}\left(j-\frac{1}{2}\right)(\hat{S}^z_j-S)\right),
   \quad (l:\text{integer})
\label{eq:twisting-operator-link}
\end{align}
and then we obtain
\begin{align}
\hat{P}_{\rm link}\hat{U}^{\rm twl}_{2\pi l}=\hat{U}^{\rm twl}_{-2\pi l}\hat{P}_{\rm link},\\
\label{pandtw2}\hat{U}^y_{\pi}\hat{U}^{\rm twl}_{2\pi
 l}=\hat{U}^{\rm twl}_{-2\pi l}\hat{U}^y_{\pi}.
\end{align}
Combining these relations, below two equations are proven\cite{Affleck-Lieb-1986,Nomura-Morishige-Isoyama-2015}:
\begin{align}
&\label{eq:elsmpu1}(\hat{P}_{\rm site}\hat{U}^y_{\pi})\hat{U}^{\rm tw}_{2\pi l}
 =(-1)^{2S l}\hat{U}^{\rm tw}_{2\pi l}(\hat{P}_{\rm site}\hat{U}^y_{\pi}),
 \\ &(\hat{P}_{\rm link}\hat{U}^y_{\pi})\hat{U}^{\rm twl}_{2\pi l}
 =\hat{U}^{\rm twl}_{2\pi l}(\hat{P}_{\rm link}\hat{U}^y_{\pi}).
 \label{eq:elsmpu2}
\end{align}

Usually, the relation between $\hat{U}^{\rm tw}_{2\pi l}$ and
$\hat{U}^{\rm twl}_{2\pi l}$ is not simple, however for the $S^z_T$=0
case it becomes
 \begin{align}
  \hat{U}^{\rm twl}_{2\pi l}|S^z_T=0\rangle
  &= \exp\left(\frac{\pi i l}{L}(\hat{S}^{z}_{T}-SL)\right)
  \hat{U}^{\rm tw}_{2\pi l}|S^z_T=0\rangle
  \notag \\
  &=(-i)^{2Sl}\hat{U}^{\rm tw}_{2\pi l}|S^z_T=0\rangle.
%  \qquad (\text{for} \; S^z_T=0 )
 \label{eq:twlandtw}
 \end{align}

\section{Discrete symmetries and the LSM theorem
\label{section:Discrete symmeties and LSM theorem}
}

\subsection{Set of the spin reversal operators' eigenvalues}

We consider the Hamiltonian with spin reversal symmetries
($[\hat{U}^x_{\pi},\hat{H}]=[\hat{U}^y_{\pi},\hat{H}]=[\hat{U}^z_{\pi},\hat{H}]=0$)
like Eq. (\ref{eq:xyz}).
In this case, one can choose a simultaneous eigenstate for the Hamiltonian
and spin reversal operators.
There is another restriction for the spin reversal operators
(see Appendix \ref{appendix:Spin-reversal-Klein}):
\begin{align}
 \hat{U}^x_{\pi}\hat{U}^y_{\pi}\hat{U}^z_{\pi} =1.
 \label{eq:Klein}
\end{align}
Therefore, the set of spin reversal eigenvalues should be 
\begin{align}
 (\hat{U}^x_{\pi},\hat{U}^y_{\pi},\hat{U}^z_{\pi})=(1,1,1),(-1,-1,1),(-1,1,-1),(1,-1,-1).
 \label{4cases}
\end{align}
In fact, since the spin reversal operators
$\hat{U}^x_{\pi},\hat{U}^y_{\pi},\hat{U}^z_{\pi}$ satisfy
($\hat{U}^x_{\pi})^2=(\hat{U}^y_{\pi})^2=(\hat{U}^z_{\pi})^2=\hat{1}$
and Eq. (\ref{eq:Klein}), the set 
\{$\hat{1},\hat{U}^x_{\pi},\hat{U}^y_{\pi},\hat{U}^z_{\pi}$\} form the
Klein four-group, which is isomorphic to the $Z_2 \times Z_2$ group.

\subsection{Expression of the states with wave number $q=\pi$ by
  twisting operators}
  
In this subsection, 
by using the twisting operators, we construct the
spin reversal eigenstates.
In the subspace $q=\pi$, we consider
the four states
\begin{align}
 \frac{i}{2}(\hat{U}^{\rm tw}_{2\pi}-\hat{U}^{\rm  tw}_{-2\pi})
 |\Psi_g\rangle
 &\equiv |\Psi_{\rm DL}\rangle,
\label{eq:d1}
 \\
 \frac{1}{2}(\hat{U}^{\rm tw}_{2\pi}+\hat{U}^{\rm tw}_{-2\pi})
 |\Psi_g\rangle
 &\equiv |\Psi_{\rm NL}\rangle,
\label{eq:n1}
 \\
 \frac{1}{2}(\hat{U}_{\pi/2}^{x}+\hat{U}_{-\pi/2}^{x})
 |\Psi_{\rm NL}\rangle
 &\equiv |\Psi_{\rm DB1}\rangle,
\label{eq:n2}
 \\
 \frac{1}{2}(\hat{U}_{\pi/2}^{y}+\hat{U}_{-\pi/2}^{y})
|\Psi_{\rm NL}\rangle
 &\equiv |\Psi_{\rm DB2}\rangle,
\label{eq:n3}
\end{align}
where the phase factors are determined so that the matrix components of the
states become real under the $\hat{S}^{z}_{j}$ diagonal representation.

At first, we can calculate the spin reversal eigenvalues of
$|\Psi_{\rm DL}\rangle$ as ($1,1,1$) and
$|\Psi_{\rm NL}\rangle$ as ($-1,-1,1$), 
by using Eqs. (\ref{uytw}) and (\ref{eq:Klein})
and 
$[\hat{S}_T^z,\hat{U}^{\rm tw}_{2\pi l}]=0$.

Next,
using the relation
$
(\hat{U}_{\pi/2}^{x})^{\dagger}\hat{U}^{y}_{\pi}\hat{U}_{\pi/2}^{x}
=(\hat{U}^{z}_{\pi} )^{\dagger}=\hat{U}^{z}_{\pi}
$
and Eq.(\ref{eq:n1}), we can derive the
eigenvalue of $\hat{U}^{y}_{\pi}$ for the state
$|\Psi_{\rm DB1}\rangle$
from the calculation
\begin{align}
 \hat{U}^{y}_{\pi}\hat{U}_{\pi/2}^{x}
 |\Psi_{NL}\rangle
 =\hat{U}_{\pi/2}^{x}\hat{U}^{z}_{\pi}
|\Psi_{NL}\rangle
 =\hat{U}_{\pi/2}^{x} |\Psi_{NL}\rangle.
\end{align}
Similarly, using the relations
$(\hat{U}_{\pi/2}^{x})^{\dagger}\hat{U}^{z}_{\pi}\hat{U}_{\pi/2}^{x}=\hat{U}^{y}_{\pi},(\hat{U}_{\pi/2}^{x})^{\dagger}\hat{U}^{x}_{\pi}\hat{U}_{\pi/2}^{x}=\hat{U}^{x}_{\pi}$, the state
$|\Psi_{\rm DB1}\rangle$
has
eigenvalues
$(\hat{U}^x_{\pi},\hat{U}^y_{\pi},\hat{U}^z_{\pi})=(-1,1,-1)$.

In summary, the four states (\ref{eq:d1}) - (\ref{eq:n3}) have the
following eigenvalues:
\begin{align}
&  \hat{U}^x_{\pi}|\Psi_{\rm DL}\rangle =|\Psi_{\rm DL}\rangle,
&& \hat{U}^y_{\pi}|\Psi_{\rm DL}\rangle =|\Psi_{\rm DL}\rangle,
&&  \hat{U}^z_{\pi}|\Psi_{\rm DL}\rangle =|\Psi_{\rm DL}\rangle,
\label{eq:d1u}
 \\
&  \hat{U}^x_{\pi}|\Psi_{\rm NL}\rangle =-|\Psi_{\rm NL}\rangle,
&& \hat{U}^y_{\pi}|\Psi_{\rm NL}\rangle =-|\Psi_{\rm NL}\rangle,
&& \hat{U}^z_{\pi}|\Psi_{\rm NL}\rangle =|\Psi_{\rm NL}\rangle,
 \label{eq:n1u}
 \\
&  \hat{U}^x_{\pi}|\Psi_{\rm DB1}\rangle =-|\Psi_{\rm DB1}\rangle,
&& \hat{U}^y_{\pi}|\Psi_{\rm DB1}\rangle =|\Psi_{\rm DB1}\rangle,
&& \hat{U}^z_{\pi}|\Psi_{\rm DB1}\rangle =-|\Psi_{\rm DB1}\rangle
 \label{eq:n2u},
 \\
&  \hat{U}^x_{\pi}|\Psi_{\rm DB2}\rangle =|\Psi_{\rm DB2}\rangle,
&& \hat{U}^y_{\pi}|\Psi_{\rm DB2}\rangle =-|\Psi_{\rm DB2}\rangle,
&& \hat{U}^z_{\pi}|\Psi_{\rm DB2}\rangle =-|\Psi_{\rm DB2}\rangle
 \label{eq:n3u}.
\end{align}

\subsection{Discrete symmetries of the DL and NL states}

In this subsection, we discuss the space inversion and time reversal
symmetries of these states.

First, we consider $|\Psi_{\rm DL}\rangle$ of Eq. (\ref{eq:d1}).
By using Eqs. (\ref{pandtw1}) - (\ref{pandtw2}) and (\ref{eq:twlandtw}),
we obtain
\begin{align}
\hat{P}_{\rm site}|\Psi_{\rm DL}\rangle &=-|\Psi_{\rm DL}\rangle,
 \label{spinvdimer1}
 \\
 \hat{P}_{\rm link}|\Psi_{\rm DL}\rangle &=|\Psi_{\rm DL}\rangle.
\label{spinvdimer2}
\end{align}
For the time reversal symmetry,
by using properties
(\ref{eq:Time-reversal-operator-1}), (\ref{eq:Time-reversal-operator-2}),
and $\hat{U}^{\rm tw}_{2\pi l}\hat{K}
= (-1)^{2 S l}\hat{K}\hat{U}^{\rm tw}_{2\pi l}$
(see Appendix \ref{appendix:Time-reversal-operator}),
for half-integer $S$,
we obtain
\begin{equation}
 \hat{K}|\Psi_{\rm DL}\rangle=|\Psi_{\rm DL}\rangle.
  \label{eq:DL-time-reversal}
\end{equation}
Therefore, $|\Psi_{\rm DL}\rangle$ corresponds to
$|\Psi_{\rm dimer-like}\rangle$
of Eq. (\ref{dimerlikeqpi}).

Similarly,
by using Eqs. (\ref{pandtw1}) - (\ref{pandtw2}) and
(\ref{eq:twlandtw}),
we show the state $|\Psi_{\rm NL}\rangle$ of Eq. (\ref{eq:n1}) satisfies
the relations below in half-integer spin:
\begin{align}
\hat{P}_{\rm site}|\Psi_{\rm NL}\rangle &=|\Psi_{\rm NL}\rangle,\\
\hat{P}_{\rm link}|\Psi_{\rm NL}\rangle &=-|\Psi_{\rm NL}\rangle,\\
\hat{K}|\Psi_{\rm NL}\rangle &=-|\Psi_{\rm NL}\rangle.
\end{align}
 Therefore, $|\Psi_{\rm NL}\rangle$ corresponds to
$|\Psi_{\rm N\acute{e}el-like}\rangle$  of Eq.(\ref{neellikeqpi}).

The rest of the $q=\pi$ states, 
i.e. $|\Psi_{\rm DB1,DB2}\rangle$,
have the symmetries
\begin{align}
\hat{P}_{\rm site}|\Psi_{\rm DB1,DB2}\rangle=|\Psi_{\rm DB1,DB2}\rangle,\\
\hat{P}_{\rm link}|\Psi_{\rm DB1,DB2}\rangle=-|\Psi_{\rm DB1,DB2}\rangle,
\end{align}
because of $[\hat{P}_{\rm
site,link},\hat{U}_{\pi/2}^{x}]=[\hat{P}_{\rm site,link},\hat{U}_{\pi/2}^{y}]=0$.

\subsection{The $q=\pi$ states with U(1) symmetry}

In the Hamiltonian with U(1) symmetry,
the energy expectation for the state (\ref{eq:n2})
is equal to that for (\ref{eq:n3}).
Using
$(\hat{U}^z_{\pi/2})^{\dagger}\hat{H}\hat{U}^z_{\pi/2}=\hat{H}$
and the relationship
$
\hat{U}^z_{\pi/2}\hat{U}^x_{\pi/2}(\hat{U}^z_{\pi/2})^{\dagger}=\hat{U}^y_{\pi/2},
$
we show
\begin{align}
 \langle \Psi_{\rm DB1}|\hat{H}|\Psi_{\rm DB1}\rangle
 &=\langle
 \Psi_{\rm NL}|(\hat{U}_{\pi/2}^{x})^{\dagger}\hat{H}\hat{U}_{\pi/2}^{x}|\Psi_{\rm NL}\rangle
 \notag \\
 &=\langle
 \Psi_{\rm NL}|(\hat{U}_{\pi/2}^{x})^{\dagger}(\hat{U}^z_{\pi/2})^{\dagger}\hat{H}\hat{U}^z_{\pi/2}\hat{U}_{\pi/2}^{x}|\Psi_{\rm NL}\rangle
 \notag\\
 &=\langle
 \Psi_{\rm NL}|(\hat{U}_{\pi/2}^{y})^{\dagger}\hat{H}\hat{U}_{\pi/2}^{y}|\Psi_{\rm NL}\rangle
 \notag \\
 &=\langle
 \Psi_{\rm DB2}|\hat{H}|\Psi_{\rm DB2}\rangle.
 \label{degeneracy}
\end{align}

Secondly, Eq. (\ref{eq:n2u}) ($\hat{U}^{z}_{\pi} | \Psi_{\rm DB1}\rangle = - |
\Psi_{\rm DB1}\rangle $)
means that the  components $S^{z}_{T} \neq 0$
(more strictly $S^{z}_{T} $ odd)
should be contained in the state $| \Psi_{\rm DB1}\rangle$.
The same statement is valid for  $| \Psi_{\rm DB2}\rangle$.
Therefore, from the extended LSM theorem \cite{Nomura-Morishige-Isoyama-2015},
these two states must be part of the continuous energy spectra for
the wave number $q$.

On the other hand, states (\ref{eq:d1u}) and (\ref{eq:n1u}) have the
eigenvalue $\hat{S}^z_T=0$, and therefore they may have an energy gap. 
Note that to satisfy Eq. (\ref{eq:elsmine}), at least one of the dimer-like
state $|\Psi_{\rm DL}\rangle$ and the N\'eel-like state $|\Psi_{\rm
NL}\rangle$ must be degenerate with the ground states (
for more details, see Sect. \ref{section:normalization}).

\subsection{The ground state with SU(2) symmetry}

In the case of SU(2) symmetry, we can classify the four cases (\ref{4cases}) as
\begin{align}
\mbox{\ singlet\ }&:&(\hat{U}^x_{\pi},\hat{U}^y_{\pi},\hat{U}^z_{\pi})=(1,1,1)\\
\mbox{\ triplet\ }&:&(\hat{U}^x_{\pi},\hat{U}^y_{\pi},\hat{U}^z_{\pi})=(-1,-1,1),(-1,1,-1),(1,-1,-1)
\end{align}
where the triplet states are degenerate in energy. 
This can be proven by using a similar argument to Eq. (\ref{degeneracy}).

It is proven that the triplet state has
a continuous energy spectrum versus wave number by using extended LSM
 theorem.
 It means that only the
 $(\hat{U}^x_{\pi},\hat{U}^y_{\pi},\hat{U}^z_{\pi})=(1,1,1)$ state can
 have an energy gap.

In summary, there are two possibilities for energy spectra for the SU(2)
symmetric case. One is that the energy spectrum is
gapless. The other is that the energy gap exists between the
dimer-like ground states and the continuous energy spectra (i.e., the
ground states are degenerate at $q=0,\pi$, and the $q=\pi$ state must
satisfy the relations (\ref{eq:d1u}), (\ref{spinvdimer1}), (\ref{spinvdimer2}), and (\ref{eq:DL-time-reversal})).

 \section{Normalization of the DL and NL states
\label{section:normalization}
 }

As we have discussed in the previous section, the dimer-like state
(\ref{eq:d1}) and N\'{e}el-like state (\ref{eq:n1}) are expressed by
twisting operators. By the way, the states $\hat{U}^{\rm
tw}_{\pm2\pi}|\Psi_{g}\rangle$ have wave number $q=\pi$, and the energy
expectation value of them is degenerate with the ground state in $O(1/L)$
because of the LSM theorem. So, one may think that both the dimer-like
and the N\'eel-like states should be always degenerate with the
$q=0$ ground state. However it seems unnatural. The answer to this paradox is in
the normalization of the states. In fact, the dimer-like and N\'{e}el-like
states are not normalized, although the component operator is
unitary. If the expectation of one of the states is zero
in the infinite limit ($L \rightarrow \infty$),
it may have an energy gap without breaking the inequality of the LSM
theorem.
 
\begin{table}[htb]
  \begin{center}
  \begin{tabular}{|c|c|c|c|c|} \hline
     & $N_{\rm DL}$ & $N_{\rm NL}$ & $G_{\rm DL}$ & $G_{\rm NL}$\\ \hline 
    case1 & Finite & Finite & 0& 0\\
    case2 & 1 & 0 & 0&May be gapped \\
    case3 & 0 & 1 & May be gapped &0 \\ \hline 
  \end{tabular}
  \caption{Possible relations of the dimer-like energy $G_{\rm DL}$
   (\ref{eq:g111}) and
the N\'eel-like energy $G_{\rm NL}$ (\ref{eq:g-1-11}),
and norms for dimer-like $N_{\rm DL}$  (\ref{eq:norma}) and
N\'eel-like $N_{\rm NL}$ (\ref{eq:normb}),
in the infinite limit ($L \rightarrow \infty$).}
  \label{table:abg}
\end{center}
\end{table}
We classify these possibilities as in Table \ref{table:abg},
where $N_{\rm DL},N_{\rm NL}$ are given by
\begin{align}
 N_{\rm DL}\equiv\langle\Psi_{\rm DL}|\Psi_{\rm DL}\rangle \ge 0,
 \label{eq:norma} \\
 N_{\rm NL}\equiv\langle\Psi_{\rm NL}|\Psi_{\rm NL}\rangle \ge 0
 \label{eq:normb},
\end{align}
and $G_{\rm DL}$ and $G_{\rm NL}$ are defined as
\begin{align}
G_{\rm DL}\equiv
 N_{\rm DL}^{-1}\langle\Psi_{\rm DL}|\hat{H}|\Psi_{\rm DL}\rangle-\langle\Psi_{g}|\hat{H}|\Psi_{g}\rangle,
\label{eq:g111} \\
G_{\rm NL}\equiv N_{\rm NL}^{-1}\langle\Psi_{\rm NL}|\hat{H}|\Psi_{\rm NL}\rangle-\langle\Psi_{g}|\hat{H}|\Psi_{g}\rangle.  
\label{eq:g-1-11}
\end{align}
Then $|\Psi_{\rm DL}\rangle$ and $|\Psi_{\rm NL}\rangle$ are candidates
of the ground state with $q=\pi$.
We can distinguish these two states as
\begin{align}
 \hat{P}_{\rm site}|\Psi_{\rm DL}\rangle &=-|\Psi_{\rm DL}\rangle,
 &
 \hat{P}_{\rm link}|\Psi_{\rm DL}\rangle &=|\Psi_{\rm DL}\rangle,
&
\hat{U}^y_{\pi}|\Psi_{\rm DL}\rangle &=|\Psi_{\rm DL}\rangle,
\\
 \hat{P}_{\rm site}|\Psi_{\rm NL}\rangle &=|\Psi_{\rm NL}\rangle,
 &
 \hat{P}_{\rm link}|\Psi_{\rm NL}\rangle &=-|\Psi_{\rm NL}\rangle,
 &
\hat{U}^y_{\pi}|\Psi_{\rm NL}\rangle &=-|\Psi_{\rm NL}\rangle.
\end{align}

Now, we will explain the reason for the classification in Table
\ref{table:abg}.
Using Eqs. (\ref{eq:d1}) and (\ref{eq:n1}), we obtain
\begin{align}
 |\Psi_{\rm NL}\rangle-i|\Psi_{\rm DL}\rangle=\hat{U}^{\rm tw}_{2\pi}|\Psi_{g}\rangle,
 \label{eq:norm}
\end{align}
where the right-hand side of Eq. (\ref{eq:norm}) is normalized ($\langle\Psi_{g}|(\hat{U}^{\rm tw}_{2\pi})^{\dagger}\hat{U}^{\rm tw}_{2\pi}|\Psi_{g}\rangle=1$).
Since $\hat{U}_{\pi}^{y}|\Psi_{\rm DL}\rangle=|\Psi_{\rm DL}\rangle$ and
$\hat{U}_{\pi}^{y}|\Psi_{\rm NL}\rangle=-|\Psi_{\rm NL}\rangle$,
the next relation is derived:
\begin{align}
 \langle\Psi_{\rm DL}|\Psi_{\rm NL}\rangle
=\langle\Psi_{\rm DL}|\hat{U}_{\pi}^{y}|\Psi_{\rm NL}\rangle
=-\langle\Psi_{\rm DL}|\Psi_{\rm NL}\rangle=0.
\end{align}
Therefore the norm of the left-hand side of Eq. (\ref{eq:norm}) is
\begin{align}
 (\langle\Psi_{\rm NL}|+i\langle\Psi_{\rm DL}|)(|\Psi_{\rm NL}\rangle-i|\Psi_{\rm DL}\rangle)
 =N_{\rm DL}+N_{\rm NL}=1.
 \label{eq:norm1}
\end{align}
Next, because of extended LSM theorem \cite{Nomura-Morishige-Isoyama-2015}, next inequality is satisfied:
\begin{align}
|\langle \Psi_g|(\hat{U}_{2\pi}^{\rm
 tw})^{\dagger}\hat{H}\hat{U}_{2\pi}^{\rm tw}-\hat{H}|\Psi_g \rangle|
 \leq O(1/L).
 \label{eq:elsmineq}
\end{align}
Since
\begin{equation}
 \langle\Psi_{\rm NL}|\hat{H}|\Psi_{\rm DL}\rangle=0,
  \label{eq:DL-NL-Hamiltonian-orthogonality}
\end{equation}
we obtain
\begin{align}
 \langle\Psi_g|(\hat{U}_{2\pi}^{\rm
 tw})^{\dagger}\hat{H}\hat{U}_{2\pi}^{\rm tw}|\Psi_g\rangle
 =\langle\Psi_{\rm NL}|\hat{H}|\Psi_{\rm NL}\rangle+\langle\Psi_{\rm DL}|\hat{H}|\Psi_{\rm DL}\rangle.
\label{eq:DL-NL-decomposition}
\end{align}
By using
Eq. (\ref{eq:g111})$\times N_{\rm DL} +$ Eq.(\ref{eq:g-1-11})$\times N_{\rm NL}$,
Eqs. (\ref{eq:norm1}) and 
(\ref{eq:DL-NL-decomposition}),
then Eq. (\ref{eq:elsmineq}) is expressed as
\begin{align}
 |N_{\rm DL} G_{\rm DL}+N_{\rm NL} G_{\rm NL}|\leq O(1/L).
 \label{eq:normresult}
\end{align}
By Eq. (\ref{eq:normresult}), we can prove Table \ref{table:abg}.

Finally, in this section we have used the symmetric property of the Hamiltonian
only at Eqs. (\ref{eq:elsmineq}) and (\ref{eq:DL-NL-Hamiltonian-orthogonality}).
For Eq. (\ref{eq:elsmineq}),
the U(1) and the translation symmetries of the Hamiltonian are
required.
For Eq. (\ref{eq:DL-NL-Hamiltonian-orthogonality}), the discrete symmetries are
required.

\section{Partial discrete symmetry
\label{section:partial-discrete-symmetry}
}

In the previous sections,
in addition to the translational and U(1) symmetries,
we have assumed the full discrete symmetries,
i.e., spin reversal {\em and} space inversion
{\em and} time reversal symmetries.
However, as we will discuss in this section, 
with a part of the discrete symmetries
(spin reversal {\em or} space inversion {\em or} time reversal),
some of the results, including Table \ref{table:abg} in Sect. \ref{section:normalization},
hold.

\subsection{Spin reversal symmetry only}

We consider the interaction with spin reversal symmetry,
whereas space inversion symmetry and time reversal symmetry are broken:
$[\hat{H},\hat{U}^{y}_{\pi} ]=0$ but
$[\hat{H},\hat{P}] \neq 0, [\hat{H},\hat{K}] \neq 0$.
An example is
\begin{equation}
 \hat{H}_{DM} = \sum_{j} \bm{S}_{j}\cdot (\bm{S}_{j+1}\times
  \bm{S}_{j+2}).
\end{equation}

In this case, one of the ground states should have the properties
\begin{align}
\hat{U}^{y}_{\pi} | \Psi_{g} \rangle = \hat{U}^{z}_{\pi} | \Psi_{g} \rangle = | \Psi_{g} \rangle,
 \hat{S}^{z}_{T} |\Psi_{g} \rangle =0,
\end{align}
whereas $\hat{P}| \Psi_{g} \rangle$
and $\hat{K} | \Psi_{g} \rangle$
are not well defined.
Although there is the possibility of a nonzero wave number $q\neq 0$
ground state,
the wave number of $|\Psi_{\rm DL}\rangle$ or
$|\Psi_{\rm NL}\rangle$
is different from that of the  $| \Psi_{g} \rangle$,
and therefore
$\langle \Psi_{\rm DL} |\Psi_{g} \rangle
= \langle \Psi_{\rm NL} | \Psi_{g} \rangle =0$. 

Using the spin reversal symmetry, one can distinguish
the dimer-like state $|\Psi_{\rm DL}\rangle$
and the N\'eel-like state $|\Psi_{\rm NL}\rangle$:
\begin{align}
 \hat{U}^y_{\pi}|\Psi_{\rm DL}\rangle=|\Psi_{\rm DL}\rangle,
\quad
\hat{U}^y_{\pi}|\Psi_{\rm NL}\rangle=-|\Psi_{\rm NL}\rangle.
\end{align}
Since
$[\hat{H},\hat{U}^{y}_{\pi} ]=0$, we show
\begin{equation}
 \langle\Psi_{\rm NL}|\hat{H}|\Psi_{\rm DL}\rangle=0.
\end{equation}
Therefore, we can obtain the same result
as Table \ref{table:abg}.

  \subsection{Space inversion symmetry only
\label{section:space-inversion-symmetry}
  }

Next
we consider the case with space inversion symmetry,
whereas the spin reversal symmetry
and time reversal symmetry are broken,
i.e.,
$[\hat{H},\hat{P}] = 0$ but
$[\hat{H},\hat{U}^{y}_{\pi} ]\neq 0,
[\hat{H},\hat{K}]\neq 0$.

In this case, one of the ground states should have the properties
\begin{align}
\hat{U}_{\rm trl}|\Psi_g\rangle=|\Psi_g\rangle,
%\hat{U}^{z}_{\pi} | \Psi_{g} \rangle = | \Psi_{g} \rangle,
\end{align}
whereas
$\hat{U}^{y}_{\pi}| \Psi_{g} \rangle$ and
$\hat{K} | \Psi_{g} \rangle$
are not well defined,
which means the possibility of a nonzero magnetization.
About space inversions, we assume either
\begin{align}
 \hat{P}_{\rm site} | \Psi_{g} \rangle =| \Psi_{g} \rangle
\end{align}
or
\begin{align}
 \hat{P}_{\rm link} | \Psi_{g} \rangle = | \Psi_{g} \rangle.
\end{align}

In Sect. \ref{section:normalization},
we saw that the $q=0,\pi$ wave number states are important,
thus we consider the following rational magnetization cases:
\begin{equation}
   \hat{S}^{z}_{T} |\Psi_{g}\rangle
    = \left(S-\frac{m}{2n} \right)L |\Psi_{g}\rangle
\label{eq:rational-magnetization-dimer}
\end{equation}
($m,n$ are integers such that $m$ and $2n$ are coprime,
independent of $L$),
where the lowest spectrum has $q\rightarrow q+\pi/n$ nontrivial
periodicity:
\begin{align}
  |E(S^{z}_{T}; q) - E(S^{z}_{T}; q+\pi/n) | \le O(1/L).
 \end{align}
  In this case, from the next relation (see Appendix \ref{appendix:Translation-twisting-operator}),
\begin{equation}
   \hat{U}_{\rm trl} \hat{U}^{\rm tw}_{\pm 2\pi n} |\Psi_{g}\rangle
    = - \hat{U}^{\rm tw}_{\pm 2\pi n}  |\Psi_{g}\rangle,
\end{equation}
the states 
$\hat{U}^{\rm tw}_{\pm 2\pi n}|\Psi_{g}\rangle$
have wave number $q=\pi$.
Note that the system size $L$ must satisfy
the condition that $L/(2n)$ is integer.

\subsubsection{Site space inversion symmetry}
  
  In the $[\hat{H},\hat{P}_{\rm site}] = 0$ case,
  we should redefine states
  (\ref{eq:d1}) and (\ref{eq:n1}) as
\begin{align}
   \frac{i}{2}(\hat{U}^{\rm tw}_{2\pi n}-\hat{U}^{\rm tw}_{-2\pi n})|\Psi_g\rangle\equiv|\Psi_{\rm DL'}\rangle,
\label{eq:d12}
   \\
   \frac{1}{2}(\hat{U}^{\rm tw}_{2\pi n}+\hat{U}^{\rm tw}_{-2\pi n})|\Psi_g\rangle\equiv|\Psi_{\rm NL' }\rangle.
   \label{eq:n12}
\end{align}

  Using the site space inversion symmetry,
one can distinguish
the dimer-like state $|\Psi_{\rm DL'}\rangle$
and the N\'eel-like state $|\Psi_{\rm NL'}\rangle$:
\begin{align}
 \hat{P}_{\rm site}|\Psi_{\rm DL'}\rangle= -|\Psi_{\rm DL'}\rangle,
\quad
\hat{P}_{\rm site}|\Psi_{\rm NL'}\rangle= |\Psi_{\rm NL'}\rangle.
\end{align}
Since
$[\hat{H},\hat{P}_{\rm site}]=0$, we show
\begin{equation}
 \langle\Psi_{\rm NL'}|\hat{H}|\Psi_{\rm DL'}\rangle=0.
\end{equation}
Therefore, we can obtain the same result
as Table \ref{table:abg}.

\subsubsection{Link space inversion symmetry}
In the $[\hat{H},\hat{P}_{\rm link}] =0$ case,
we should redefine states 
(\ref{eq:d1}) and (\ref{eq:n1}) as
\begin{align}
   \frac{1}{2}(\hat{U}^{\rm twl}_{2\pi n}+\hat{U}^{\rm twl}_{-2\pi n})|\Psi_g\rangle\equiv|\Psi_{\rm DL''}\rangle,
\label{eq:d13}
   \\
   \frac{i}{2}(\hat{U}^{\rm twl}_{2\pi n}-\hat{U}^{\rm twl}_{-2\pi n})|\Psi_g\rangle\equiv|\Psi_{\rm NL''}\rangle.
   \label{eq:n13}
\end{align}

Using the link space inversion symmetry,
one can distinguish
the dimer-like state $|\Psi_{\rm DL''}\rangle$
and the N\'eel-like state $|\Psi_{\rm NL''}\rangle$:
\begin{align}
 \hat{P}_{\rm link}|\Psi_{\rm DL''}\rangle= |\Psi_{\rm DL''}\rangle,
\quad
\hat{P}_{\rm link}|\Psi_{\rm NL''}\rangle= - |\Psi_{\rm NL''}\rangle.
\end{align}
Since
$[\hat{H},\hat{P}_{\rm link}]=0$, we show
\begin{equation}
 \langle\Psi_{\rm NL''}|\hat{H}|\Psi_{\rm DL''}\rangle=0.
\end{equation}
Therefore, we can obtain the same result
as Table \ref{table:abg}.

Note that when there is spin reversal or time reversal symmetry,
Eqs. (\ref{eq:d13}) and (\ref{eq:n13}) satisfy
\begin{align}
 \hat{U}^{y}_{\pi} |\Psi_{\rm DL''}\rangle = |\Psi_{\rm DL''}\rangle,
 \quad
 \hat{U}^{y}_{\pi} |\Psi_{\rm NL''}\rangle = -|\Psi_{\rm NL''}\rangle
\end{align}
or
\begin{align}
 \hat{K} |\Psi_{\rm DL''}\rangle = |\Psi_{\rm DL''}\rangle,
 \quad
 \hat{K} |\Psi_{\rm NL''}\rangle = -|\Psi_{\rm NL''}\rangle.
\end{align}

\subsection{Time reversal symmetry only}

Finally we consider the interaction with time reversal symmetry,
whereas the space inversion symmetry and spin reversal symmetry are broken:
$[\hat{H},\hat{K}]=0$ but
$[\hat{H},\hat{P}] \neq 0,
[\hat{H},\hat{U}^{y}_{\pi} ] \neq 0$. 
An example is
\begin{equation}
 \hat{H}_{DM} = \sum_{j} (\bm{S}_{j}\times \bm{S}_{j+1})^{z}.
\end{equation}

In this case, one of the ground states should have the properties
\begin{align}
\hat{K} | \Psi_{g} \rangle = | \Psi_{g} \rangle,
 \hat{S}^{z}_{T} |\Psi_{g} \rangle =0,
\end{align}
whereas
$\hat{P}| \Psi_{g} \rangle$
and
$\hat{U}^{y}_{\pi}| \Psi_{g} \rangle$
are not well defined.
Although there is a possibility of the $q\neq 0$
ground state,
the wave number of $|\Psi_{\rm DL}\rangle$ or
$|\Psi_{\rm NL}\rangle$
is different from that of the  $| \Psi_{g} \rangle$,
and therefore
$\langle \Psi_{\rm DL} |\Psi_{g} \rangle
= \langle \Psi_{\rm NL}|\Psi_{g} \rangle =0$.

Using the time reversal symmetry, one can distinguish
the dimer-like state $|\Psi_{\rm DL}\rangle$
and the N\'eel-like state $|\Psi_{\rm NL}\rangle$
as
\begin{align}
 \hat{K} |\Psi_{\rm DL}\rangle=|\Psi_{\rm DL}\rangle,
\quad
\hat{K} |\Psi_{\rm NL}\rangle=-|\Psi_{\rm NL}\rangle.
\end{align}
Since
$[\hat{H},\hat{K} ]=0$, we show
\begin{equation}
 \langle\Psi_{\rm NL}|\hat{H}|\Psi_{\rm DL}\rangle=0.
\end{equation}
Therefore, we can obtain the same result
as Table \ref{table:abg}.

 \section{Examples
\label{section:examples}
 }
 
Since we have used symmetries except short-range interaction,
we can apply our consideration for various models.

  \subsection{$S=1/2$ spin chain with next-nearest-neighbor interaction
\label{section:S=1/2-NNN-XXZ}
  }

 \subsubsection{XXZ case}
We consider the $S=1/2$ XXZ spin chain with nearest- and
next-nearest-neighbor (NNN) interactions:
\begin{align}
 H = J_1 \left(\sum_{j}  \hat{S}^{x}_{j}\hat{S}^{x}_{j+1}
  +\hat{S}^{y}_{j}\hat{S}^{y}_{j+1} +\Delta
  \hat{S}^{z}_{j}\hat{S}^{z}_{j+1} \right)
 + J_2 \left(\sum_{j}  \hat{S}^{x}_{j}\hat{S}^{x}_{j+2}
  +\hat{S}^{y}_{j}\hat{S}^{y}_{j+2} +\Delta \hat{S}^{z}_{j}\hat{S}^{z}_{j+2} \right)
\label{eq:S=1/2-NNN-XXZ-chain}
\end{align}
At $J_2=0$, there is an exact solution by Bethe Ansatz,
where $\Delta \le 1$ gapless, and 
$\Delta > 1$ is the gapped N\'eel ordered phase.
At $J_2/J_1=1/2$, this model was solved by
\cite{Majumdar-Ghosh-1969,Majumdar-Ghosh-1970} and \cite{t2},
and it is known that the model has the twofold
degenerate dimer ground states as Eq. (\ref{neelanddimer2}).

In the region ($J_1>0, \: J_2/J_1\le 1/2, \: \Delta>0$)
\cite{Nomura-Okamoto-1993,Nomura-Okamoto-1994}
there are three phases: the N\'eel, dimer and spin-fluid (XY) phases.
The N\'eel phase corresponds to case 3 of Table  \ref{table:abg} in
Sect. \ref{section:normalization},
the dimer phase corresponds to case 2,
and finally the spin-fluid phase corresponds to case 1
(where the doublet states
($S^{z}_{T}=\pm 1, q=\pi,\hat{P}_{\rm site}=1,\hat{P}_{\rm link}=-1$)
are the lowest energy excitation).
The N\'eel and dimer phases are gapped, whereas the spin-fluid phase
is gapless. 

About the phase boundary,
between the N\'eel phase and the dimer phase 
there is a second-order transition line (Gaussian
line). 
The Gaussian line is characterized by the energy level crossing of the
N\'eel-like state
($S^{z}_{T}=0, q=\pi,\hat{U}^{y}_{\pi}=-1,
\hat{P}_{\rm site}=1,\hat{P}_{\rm link}=-1,\hat{K}=-1$)
and the dimer-like state
($S^{z}_{T}=0, q=\pi,\hat{U}^{y}_{\pi}=1,
\hat{P}_{\rm site}=-1,\hat{P}_{\rm link} = 1, \hat{K}=1$).

The phase boundary between the spin-fluid phase and the N\'eel phase,
which lies on the $\Delta =1, J_2/J_1 \le 0.2411$,
and the phase boundary between
the spin-fluid and the dimer phases,
are of the Berezinskii-Kosterlitz-Thouless (BKT) type.
The BKT phase boundaries are determined  by the energy level crossing
of the doublet states and the N\'eel-like state,
or the doublet states and the dimer-like state (the level spectroscopy
method,
see \cite{Nomura-Okamoto-1994}
and \cite{Nomura-Kitazawa-1998}
).
The BKT multicritical point
at $\Delta =1, J_2/J_1 = 0.2411$ \cite{Okamoto-Nomura-1992}
is determined by the energy level crossing
of the triplet (the doublet states and the N\'eel-like state )
and the dimer-like state.
In summary, the consideration on symmetries in this work
gives another support for the level spectroscopy,
besides the field theory and the renormalization.

In the region ($J_1>0,\: J_2/J_1>1/2, \: \Delta>1$)
\cite{Igarashi-Tonegawa-1989},
there are the dimer phase and the (2,2) antiphase
(or up-up-down-down) state phase.
In the latter phase, there are the four ground states
$q=0,\pm\pi/2,\pi$, and above them there is an energy gap.
The phase transition between the dimer and the (2,2) antiphase state
phase is considered of the 2D Ising universality class.

 \subsubsection{Isotropic case}
 Next we consider the isotropic NNN chain,
 i.e., $\Delta=1$ in
 Eq. (\ref{eq:S=1/2-NNN-XXZ-chain}).

In the region $J_{1}<0, \: J_{2}/|J_{1}| \le 1/4$,
there is a ferromagnetic phase.

In the region $J_{1}>0, \: J_{2}/J_{1} \le 0.2411$ \cite{Okamoto-Nomura-1992},
this model is gapless at the wave number
$q=0,\pi$, and the universality class is of a Tomonaga-Luttinger type.
In the region $J_{1}>0, \: J_{2}/J_{1}>0.2411$,
there is a gapped phase with the dimer-long range order.

At $J_{1}=0, \: J_{2}>0$,
the model (\ref{eq:S=1/2-NNN-XXZ-chain}) becomes two decoupled spin chains,
thus it is gapless at the four wave numbers $q=0,\pm \pi/2, \pi$.

In the region  $J_{1}<0, \: J_{2}/|J_{1}|>1/4$ 
there has been controversial studies into whether it is gapless or gapped.
On the basis of the renormalization group,
it has been discussed that there should be a very small gap
(or very long correlation length)
\cite{Itoi-Qin-2001}.
Finally,
with careful numerical calculations, it has been shown 
that this region is gapped and it is named 
the Haldane-dimer phase\cite{Furukawa-Sato-Onoda-Furusaki-2012}.

Note that
at the point $J_{1}<0, \: J_{2}/|J_{1}|=1/4$,
the ground state of Eq. (\ref{eq:S=1/2-NNN-XXZ-chain}) with $\Delta=1$
highly degenerate\cite{Suzuki-Takano-2008},
 and there is an RVB-like state in them
\cite{Hamada-Kane-Nakagawa-Natsume-1988}.

\subsection{Spin ladder model}
\begin{figure}[h]
\begin{center}
 \includegraphics[height=3cm,angle=0]{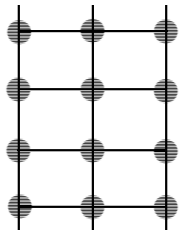}
 \caption{Three leg spin ladder.}
 \label{spinladder}
\end{center}
\end{figure}

Next, we consider the spin ladder model \cite{st1,st2} like Fig. \ref{spinladder}.
The Hamiltonian in this case is (by defining~$\hat{H}^{i,j}_{n,m}\equiv \hat{S}_{n,i}\hat{S}_{m,j}$)
\begin{align}
\notag \hat{H}=\sum^N_{n=1}[J_1\sum_{i=1}^{2}\hat{H}^{i,i+1}_{n,n}+J_2\hat{H}^{1,3}_{n,n}+J_{\perp 1}\sum_{i=1}^{3}\hat{H}^{i,i}_{n,n+1} \\\notag+J_3\sum_{i=1}^{2}\hat{H}^{i,i+1}_{n,n+1}+J_3\sum_{i=2}^{3}\hat{H}^{i,i-1}_{n,n+1}\\+J_{\perp 2}\sum_{i=1}^{3}\hat{H}^{i,i}_{n,n+2}],
\end{align}
 where $L$ is the system size of leg direction and 3 is the rung direction's one.
 
  In the $S=1/2$ three-leg ladder case,
we can obtain the same result for the $S$ half-integer spin chains: the energy
spectrum is gapless or gapped with dimer-like state. Generally speaking,
we can apply our discussion to the half-integer spin ladder with
odd leg. (So, if we know a particular system has a gap already, the system
has the dimer-like state as the ground state.)

\subsection{Spin tube model}
\begin{figure}[h]
\begin{center}
 \includegraphics[height=4cm,angle=0]{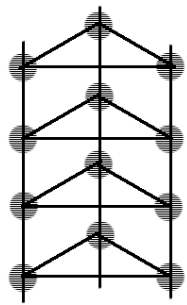}
 \caption{Three leg spin tube.}
\label{spin tube}
\end{center}

\end{figure}

The spin tube model (Fig. \ref{spin tube}) with nearest-neighbor
interaction may have an energy gap, depending on parameters.
This is similar to the previous one, but there is a difference between
the two cases in the interaction to the rung direction\cite{st1,st2}. This lattice forms the $C_{3v}$ point group for the rung direction.

The Hamiltonian is
\begin{align}
 \hat{H}=J_1\sum^L_{n=1}\sum_{i=1}^{3}\hat{H}^{i,i+1}_{n,n}+J_{\perp}\sum^L_{n=1}\sum_{i=1}^{3}\hat{H}^{i,i}_{n,n+1}.
\end{align}

\subsection{Operator for the spin tube model and the spin ladder model}
 Although the operators should be modified, we can apply our method to the spin ladder and spin tube. The modifications are
 \begin{align}
\label{sltux}\hat{U}^x_{\theta}\equiv\exp{(-i\theta \hat{S}^x_T)},~\hat{S}^x_T\equiv\sum^L_{n=1}\sum^3_{i=1}\hat{S}^x_{n,i}\quad \text{etc.},
 \end{align}
\begin{equation}
\hat{U}^{\rm tw}_{\pm2\pi}\equiv\exp{\left(\mp\frac{2\pi i}{L}\sum^L_{n=1}\sum^3_{i=1}j(\hat{S}^z_{n,i}-S)\right)},
\end{equation}
 \begin{align}
\label{sltps} \hat{P}^{\dagger}_{\rm site}\hat{S}^{x,y,z}_{n,i}\hat{P}_{\rm site}=\hat{S}^{x,y,z}_{N-n,i},\\
\label{sltpl}\hat{P}^{\dagger}_{\rm link}\hat{S}^{x,y,z}_{n,i}\hat{P}_{\rm link}=\hat{S}^{x,y,z}_{N-n+1,i}.
\end{align} 
In the half-integer spin ladder and spin tube system, by taking the rung
direction's 3 sites as a unit-cell, we can also say that the ground states
are dimer-like in the same way as the above 1D spin system. To sum
up, the energy spectrum is continuous (corresponding to the N\'{e}el-like state)
or gapped (corresponding to the dimer-like state) in the spin ladder and spin
tube as well as the pure 1D spin system. Of course, the N\'{e}el-like state
has $\hat{P}_{\rm site}=1,\hat{P}_{\rm link}=-1$ and dimer-like state
has $\hat{U}^x_{\pi}=\hat{U}^y_{\pi}=\hat{U}^z_{\pi}=1,\hat{P}_{\rm
site}=-1,\hat{P}_{\rm link}=1$ where $\hat{U}^x_{\pi},\hat{P}_{\rm
site},\hat{P}_{\rm link}$ are defined by Eqs. (\ref{sltux}),(\ref{sltps}),
and (\ref{sltpl}).

\begin{figure}[h]
\begin{center}
 \includegraphics[height=3.5cm,angle=0]{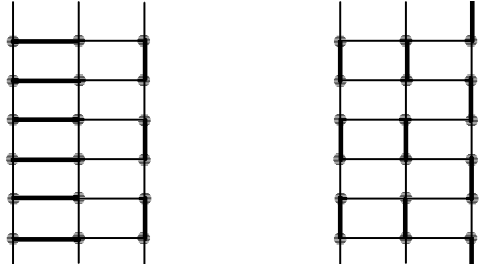}
 \caption{Spin ladder dimer-like state. Space inversion is odd, where
 space inversion is on the second site of the rung direction.}
 \label{Space inversion is odd}
\end{center}
\end{figure}

\begin{figure}[h]
\begin{center}
 \includegraphics[height=3.5cm,angle=0]{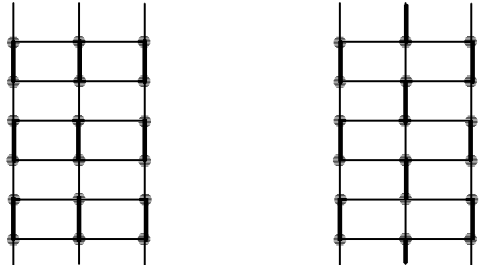}
 \caption{Space inversion is even.}
 \label{Space inversion is even}
\end{center}
\end{figure}

We point out four suitable dimer-like states in Figs. \ref{Space inversion is odd} and \ref{Space inversion is even}, and categorize with these space inversion symmetry for rung direction.

And its corresponding wave functions can be written; e.g., the left one of Fig. \ref{Space inversion is odd} is expressed by
\begin{align}
|\Psi_{dimer,l1,r1}\rangle+|\Psi_{dimer,l2,r1}\rangle+|\Psi_{dimer,l1,r2}\rangle+|\Psi_{dimer,l2,r2}\rangle : q=0\\~|\Psi_{dimer,l1,r1}\rangle-|\Psi_{dimer,l2,r1}\rangle+|\Psi_{dimer,l1,r2}\rangle-|\Psi_{dimer,l2,r2}\rangle : q = \pi\label{spinladderdimer1}
\end{align}
where
\begin{align}
 |\Psi_{dimer,l1,r1}\rangle
 &\equiv [1,1,2][1,3,4]\dots [1,L-1,L]
 \notag \\
 &\times \{2,3,1\}\{2,3,2\}\dots\{2,3,L\}
 \\
 |\Psi_{dimer,l2,r1}\rangle
 &\equiv [1,2,3][1,4,5]\dots [1,L,1]
 \notag \\
 &\times \{2,3,1\}\{2,3,2\}\dots\{2,3,L\}
 \\
 |\Psi_{dimer,l1,r2}\rangle
 &\equiv [1,1,2][1,3,4]\dots [1,L-1,L]
 \notag \\
 &\times \{2,1,1\}\{2,1,2\}\dots\{2,1,L\}
 \\
 |\Psi_{dimer,l2,r2}\rangle
 &\equiv [1,2,3][1,4,5]\dots [1,L,1]
 \notag \\
 &\times
 \{2,1,1\}\{2,1,2\}\dots\{2,1,L\}
 \\
 [i,j,k]&=\frac{1}{\sqrt{2}}(\uparrow_{j,i}\downarrow_{k,i}-\downarrow_{j,i}\uparrow_{k,i})
 \notag \\
\{i,j,k\}&=\frac{1}{\sqrt{2}}(\uparrow_{k,i}\downarrow_{k,j}-\downarrow_{k,i}\uparrow_{k,j})
\notag
\end{align}
Note $\prod_k\{i,j,k\}$ always has $\hat{U}^y_{\pi}=1,\hat{P}_{\rm
site}=1,\hat{P}_{\rm link}=1$ as eigenvalues (this space inversion is
the leg direction's one), so we can easily understand that
Eq. (\ref{spinladderdimer1}) has $\hat{U}^y_{\pi}=1,\hat{P}_{\rm
site}=-1,\hat{P}_{\rm link}=1$ with the possibility of an energy gap.

\subsection{Distorted diamond chain model}
Our discussion can be also valid in a distorted diamond chain \cite{Okamoto-Tonegawa-Kaburagi-2003}.
The Hamiltonian is
\begin{align}
\hat{H}=J_1\sum^{N/3}_{j=1}
 (\hat{S}_{3j-1}\hat{S}_{3j}+\hat{S}_{3j}\hat{S}_{3j+1})+J_2\sum^{N/3}_{j=1}\hat{S}_{3j+1}\hat{S}_{3j+2}+J_3\sum^{N/3}_{j=1}(\hat{S}_{3j-2}\hat{S}_{3j}+\hat{S}_{3j}\hat{S}_{3j+2}).
 \label{eq:distorted-diamond-chain}
\end{align}

The diamond chain has 3 sites in unit-cell like the spin ladder and spin tube.
In addition, this model has space inversion
symmetry.
(K. Okamoto, Shibaura Institute of Technology, private communicationy) 
Therefore, by the same argument, the energy spectrum is gapless or gapped with dimer-like state.

   \subsubsection{Magnetic plateaux
\label{section:magnetic-plateaux}
   }

In a nonzero magnetic field, the symmetry of
Eq. (\ref{eq:distorted-diamond-chain})
becomes from SU(2) to U(1).
In the rational magnetization $M=M_s/3,2 M_s/3$
($M_s$ is the saturation magnetization),
it is reported that magnetization plateaux appear
\cite{Okamoto-Tonegawa-Kaburagi-2003},
by using the level spectroscpoy and other methods.
The BKT phase transition at $2 M_s/3$ is a similar type
as the S=1/2 NNN XXZ model,
since the discussion of Sect. 
\ref{section:space-inversion-symmetry}
applies for the $2 M_s/3$ phase diagram.

 \section{Ground state symmetries and system size
\label{section:ground-state-symmetries}
 }

\subsection{System size $L=2n+1$}

In the case of $SL$ half-integer (i.e., $S$ half-integer and $L$ odd),
the relation of spin reversal operators
becomes
$(\hat{U}_{\pi}^x)^2=(\hat{U}_{\pi}^y)^2=(\hat{U}_{\pi}^z)^2=-1$,
and
$\hat{U}_{\pi}^x\hat{U}_{\pi}^y\hat{U}_{\pi}^z=-1$.
Therefore the sets
$\{\pm\hat{1},\pm\hat{U}_{\pi}^x,\pm\hat{U}_{\pi}^y,\pm\hat{U}_{\pi}^z\}$
form the quaternion group
(see Appendix \ref{appendix:Spin-reversal-Klein}).
The time reversal operator has a similar relation $\hat{K}^2=-1$.
Thus, the algebraic structure of the $SL$ half-integer case
is quite different from the $SL$ integer case.

Since we did not use the periodic boundary condition in the process of
the derivation in Appendix \ref{appendix:Spin-reversal-Klein},
we can apply $\hat{U}_{\pi}^x\hat{U}_{\pi}^y\hat{U}_{\pi}^z=-1$ for
the $SL$ half-integer case, 
or $\hat{U}_{\pi}^x\hat{U}_{\pi}^y\hat{U}_{\pi}^z=1$
for the $SL$  integer case, 
also in the open boundary condition. 

In the  periodic boundary condition, we cannot distinguish the two types
of space inversion (site inversion and link inversion) in the odd sites
system. On the other hand, in the open boundary condition, only the site inversion can be defined.

\subsection{System size $L=4n,4n+2$}

For the nonfrustrating case, by using an MLM-type discussion
\cite{Marshall-1955} \cite{Lieb-Mattis-1962},
one can prove that the ground state has
$q=0,\hat{U}^y_{\pi}=1,\hat{P}_{\rm link}=\hat{P}_{\rm site}=1$ in
$L=4n$,
and
$q=\pi,\hat{U}^y_{\pi}=1,\hat{P}_{\rm link}=-1,\hat{P}_{\rm site}=1$ in $L=4n+2$\cite{Barwinkel-Schmidt-Schnack-2000}.
Therefore, for $L=4n+2$,
we should choose
$|\Psi_g\rangle$ as $\hat{U}_{\rm trl}|\Psi_g\rangle=-|\Psi_g\rangle,
\hat{U}^{y}_{\pi} | \Psi_{g} \rangle = \hat{U}^{z}_{\pi} | \Psi_{g} \rangle = | \Psi_{g} \rangle,
\hat{P}_{\rm site} | \Psi_{g} \rangle = | \Psi_{g}\rangle, \hat{P}_{\rm
link} | \Psi_{g} \rangle = -| \Psi_{g}\rangle $
instead of Eq. (\ref{eq:ground-state-condition}),
but basically our discussions for $L=4n$ are valid also in $L=4n+2$.

 \section{Summary and discussions
\label{section:summary}
 }

In this paper we have investigated the system with discrete symmetries:
spin reversal, space inversion (link 
or site inversion), and time
reversal symmetries.
And when the system has U(1) or SU(2) continuous symmetry,
we can discuss the low-lying states using the LSM theorem.
In Sect. 
\ref{section:Discrete symmeties and LSM theorem}
we constructed four states with the twisting operators,
then confirmed the discrete symmetries of them.
Using the discrete symmetries,
we classify the $q=\pi$ states
into the dimer-like, the N\'eel-like, or other states.
Although superficially the LSM theorem suggests
both the dimer-like and the N\'eel-like states ($q=\pi$) are degenerate with
the $q=0$ state,
we have solved this dilemma by considering the normalization
of the dimer-like and N\'eel-like states in Sect. 
\ref{section:normalization},
where the relation of the normalization and the gap
is summarized in Table \ref{table:abg}.
In Sect. \ref{section:partial-discrete-symmetry},
we have shown that a part of the discrete symmetries,
i.e.,
either space inversion {\em or}
spin reversal {\em or} time reversal,
is enough for the above arguments.
Note that since the spin reversal or time reversal symmetry means
zero magnetization,
then the space inversion symmetry is needed to explain the magnetic
plateau
(see also the example in Sect. \ref{section:magnetic-plateaux}).

We mention the relation between the BKT type transition and the present work.
As we discussed in Sect. \ref{section:S=1/2-NNN-XXZ},
the classification of Table \ref{table:abg}
is closely related with the BKT type transition.
However, when the number of the degenerate ground states
is more than the minimum degeneracy expected from the LSM theorem,
such as the (2,2) antiphase state in Sect. \ref{section:S=1/2-NNN-XXZ},
it is not necessarily related with the BKT type transition.
Moreover one should classify whether the doublet state is gapped or gapless,
plus the classification in Table \ref{table:abg}.

There is caution on the statement that at least  
one of the ground states 
satisfies Eq. (\ref{eq:ground-state-condition})
in Sect. \ref{section:ground-state}.
This statement may remain valid
in the frustrating regions with zero magnetization,
including the Majumdar-Gosh model.
Generally, in the rational magnetization with space inversion symmetry,
a similar statement, at least 
one of the ground states 
satisfies $ \hat{P}| \Psi_{g} \rangle =| \Psi_{g} \rangle$ and $q=0$,
may be valid.
However 
the above statement is violated for the irrational magnetization with frustration
when the incommensurability occurs.
For example, in the NNN XXZ model (\ref{eq:S=1/2-NNN-XXZ-chain}),
the energy dispersion of the one spin flip from the fully aligned
state
($S^{z}_{T}=SL-1$)
is the double well form
when $J_1>0, \: J_2/J_1 >1/4$.

Finally, we comment on the symmetry protected topological phases
(SPTP)
\cite{Pollmann-Turner-Berg-Oshikawa-2010}\cite{Pollmann-Berg-Turner-Oshikawa-2012}.
Between our work and SPTP,
there is a similar point of view (distinguishing the states
by space inversion, spin reversal, and time reversal symmetry), 
however, some conditions are different (SPTP: integer spin, our work: half-integer spin).
Other differences are that SPTP theory treats only the link 
inversion as the space inversion, whereas in our theory both the site
inversion and the link inversion are considered; 
in our theory the magnetic plateaux in the rational magnetization
have been treated.

\section*{Acknowledgement}
We want to thank K. Hida,
M. Sato,
H. Katsura and K. Okamoto for giving insightful comments and suggestions.

\bigskip

\appendix

 \section{
$Z_{2} \times Z_{2}$ spin reversal symmetry 
\label{appendix:Spin-reversal-Klein}
}

[Lemma 1 \label{lemma:1} ]. 
For the $SL$ integer case,
\begin{align}
 \hat{U}^x_{\pi}\hat{U}^y_{\pi}\hat{U}^z_{\pi}
 = \hat{U}^x_{2\pi}
 = \exp{(-2\pi i \hat{S}^x_T)}
 =1.
\end{align}

\noindent
[Proof]
According to Eq. (\ref{spinrotation}), 
\begin{align}
(\hat{U}^x_{\pi/2})^{\dagger}\hat{S}^y_k\hat{U}^x_{\pi/2}=-\hat{S}^z_k.
\end{align}
Therefore, we obtain
\begin{align}
 (\hat{U}^x_{\pi/2})^{\dagger}\hat{U}^y_{\pi/2}\hat{U}^x_{\pi/2}
  =(\hat{U}^x_{\pi/2})^{\dagger}\left[\sum_{n=0}^{\infty}\frac{(-i\pi/2)^n}{n!}(S^y_T)^n\right]\hat{U}^x_{\pi/2}
 =(\hat{U}^z_{\pi/2})^{\dagger},
\end{align}
or
\begin{align}
\label{uuu2}\hat{U}^y_{\pi/2}\hat{U}^x_{\pi/2}=\hat{U}^x_{\pi/2}(\hat{U}^z_{\pi/2})^{\dagger}.
\end{align}

Similarly we get
\begin{align}
 \hat{U}^y_{\pi/2}\hat{U}^z_{\pi/2}
 =\hat{U}^x_{\pi/2}\hat{U}^y_{\pi/2}
 \label{uuu1}
\end{align}
and
\begin{align}
 (\hat{U}^z_{\pi/2})^{\dagger}\hat{U}^y_{\pi/2}\hat{U}^z_{\pi/2}
 =\hat{U}^x_{\pi/2}.
 \label{uuu3}
\end{align}

 By using Eqs. (\ref{uuu1}),(\ref{uuu2}), and (\ref{uuu3}) one after another, we obtain the product relation of spin reversal operators as
\begin{align}
 \hat{U}^x_{\pi}\hat{U}^y_{\pi}\hat{U}^z_{\pi}
 &=
 \hat{U}^x_{\pi/2}\hat{U}^x_{\pi/2}\hat{U}^y_{\pi/2}(\hat{U}^y_{\pi/2}\hat{U}^z_{\pi/2})\hat{U}^z_{\pi/2}
\notag \\
 &=\hat{U}^x_{\pi/2}\hat{U}^x_{\pi/2}(\hat{U}^y_{\pi/2}\hat{U}^x_{\pi/2})\hat{U}^y_{\pi/2}\hat{U}^z_{\pi/2}
  \notag \\
 &=\hat{U}^x_{\pi/2}\hat{U}^x_{\pi/2}\hat{U}^x_{\pi/2}[(\hat{U}^z_{\pi/2})^{\dagger}\hat{U}^y_{\pi/2}\hat{U}^z_{\pi/2}]
 \notag \\
 &=\hat{U}^x_{\pi/2}\hat{U}^x_{\pi/2}\hat{U}^x_{\pi/2}\hat{U}^x_{\pi/2}
 \notag \\
&=\hat{U}^x_{2\pi}.
\end{align}

Finally, since the eigenvalue of $\hat{S}^x_T$ is integer for
$SL$:integer,
we can prove
\begin{align}
 \hat{U}^x_{\pi}\hat{U}^y_{\pi}\hat{U}^z_{\pi}
 = \hat{U}^x_{2\pi}
 = \exp{(-2\pi i \hat{S}^x_T)}
 =1.
\end{align}
[Remarks]

(I) These spin reversal operators
$\hat{U}^x_{\pi},\hat{U}^y_{\pi},\hat{U}^z_{\pi}$ satisfy
($\hat{U}^x_{\pi})^2=(\hat{U}^y_{\pi})^2=(\hat{U}^z_{\pi})^2=\hat{1}$
and $\hat{U}^x_{\pi}\hat{U}^y_{\pi}=\hat{U}^z_{\pi}$. Therefore
\{$\hat{1},\hat{U}^x_{\pi},\hat{U}^y_{\pi},\hat{U}^z_{\pi}$\} form the
Klein four-group, which is isomorphic to the $Z_2 \times Z_2$ group.

 (II) This proof does not need the translational, space inversion,
 and time reversal symmetries. 

\bigskip 

\noindent
[Corollary 1]
For the $SL$ half-integer case,
\begin{align}
 \hat{U}^x_{\pi}\hat{U}^y_{\pi}\hat{U}^z_{\pi}
 =-1.
\end{align}
and 
$(\hat{U}_{\pi}^x)^2=(\hat{U}_{\pi}^y)^2=(\hat{U}_{\pi}^z)^2=-1$,
since each of the eigenvalues 
$\hat{S}^{x}_T, \hat{S}^{y}_T, \hat{S}^{z}_T$ is half-integer.

Therefore 
$\{\pm\hat{1},\pm\hat{U}_{\pi}^x,\pm\hat{U}_{\pi}^y,\pm\hat{U}_{\pi}^z\}$
form the quaternion group.

 \section{Translation and twisting operator
\label{appendix:Translation-twisting-operator}
 }

[Lemma 2 \label{lemma:2} ]
 \begin{align}
\hat{U}^{\rm tw}_{2\pi l} \hat{U}_{\rm trl}
&=  \hat{U}_{\rm trl}\hat{U}^{\rm tw}_{2\pi l}
 \exp\left(\frac{2 \pi i l}{L}(\hat{S}^{z}_{T}
 -SL)\right),
 \label{eq:TBC-translation-2pi}
\\
\hat{U}^{\rm twl}_{2\pi l} \hat{U}_{\rm trl}
&=  \hat{U}_{\rm trl}\hat{U}^{\rm twl}_{2\pi l}
 \exp\left(\frac{2 \pi i l}{L}(\hat{S}^{z}_{T}
 -SL)\right)
 \quad (l: \text{integer}).
  \label{eq:TBC-translation-2pi-2}
 \end{align}

\noindent 
[Proof]
  \begin{align}
\hat{U}_{\rm trl}^{\dagger} \hat{U}^{\rm tw}_{2\pi l} \hat{U}_{\rm trl}
&=\exp\left(- \frac{2\pi i l}{L}\sum_{j=1}^{L} j(\hat{S}^{z}_{j+1} -S)\right)
\notag \\
&=\exp\left(-\frac{2\pi i l}{L}\left(\sum_{j=2}^{L} (j-1)(\hat{S}^{z}_{j} -S)
+L(\hat{S}^{z}_{L+1}-S)\right)
\right)
\notag \\
&=\hat{U}^{\rm tw}_{2\pi l} \exp\left(\frac{2\pi i l}{L}(\hat{S}^{z}_{T}
 -SL)\right)\exp(- 2\pi i l(\hat{S}^{z}_{1} -S)),
 \label{eq:TBC-translation-2pi-3}
\end{align}
where we have used  $\hat{S}^z_{L+1} = \hat{S}^z_{1} $.
Since the eigenvalue of $\hat{S}^{z}_{1} -S$ is an integer,
we obtain Eq.
(\ref{eq:TBC-translation-2pi}).
Similarly, we obtain Eq. (\ref{eq:TBC-translation-2pi-2}).

 \section{Time reversal operator
\label{appendix:Time-reversal-operator}
 }

[Lemma 3 \label{lemma:3}]
 \begin{align}
\hat{U}^{x,y,z}_{\pi}\hat{K}
&= \hat{K}\hat{U}^{x,y,z}_{\pi}.
 \label{eq:TR-SR}
 \end{align}

\noindent
[Proof]
\begin{align}
\hat{K}^{\dagger}\hat{U}^{x,y,z}_{\pi}\hat{K}&=\hat{K}^{\dagger}\left(\sum_{n=0}^{\infty}[\{\pi
  i\sum^L_{j=1} (\hat{S}^{x,y,z}_j)\}^n/n!]\right)\hat{K}\notag\\
  &=\sum_{n=0}^{\infty}[\{-\pi
  i\sum^L_{j=1} (-\hat{S}^{x,y,z}_j)\}^n/n!]
  =\hat{U}^{x,y,z}_{\pi}.
\end{align}

\bigskip

\noindent
[Lemma 4 \label{lemma:4} ]
For the $SL$ integer case,
 \begin{align}
\hat{U}^{\rm tw}_{2\pi l}\hat{K}
&= (-1)^{2Sl}\hat{K}\hat{U}^{\rm tw}_{2\pi l},
 \label{eq:R-twist}
\\
\hat{U}^{\rm twl}_{2\pi l}\hat{K}
&= \hat{K}\hat{U}^{\rm twl}_{2\pi l}
  \label{eq:TR-twist2}
\qquad (l: \text{integer}).
 \end{align}

\noindent
[Proof]
\begin{align}
 \hat{K}^{\dagger}\hat{U}^{\rm tw}_{2\pi l}\hat{K}
 &=\hat{K}^{\dagger}\left(\sum_{n=0}^{\infty}[\{-\frac{2\pi
  i l}{L}\sum^L_{j=1}j(\hat{S}^z_j-S)\}^n/n!]\right)\hat{K}\notag\\
  &=\sum_{n=0}^{\infty}[\{\frac{2\pi
  i l}{L}\sum^L_{j=1}j(-\hat{S}^z_j-S)\}^n/n!]\notag\\
  &=\exp\left(-\frac{2\pi i l}{L}\sum^L_{j=1}j(\hat{S}^z_j-S)\right)
 \exp \left(-\frac{2\pi i l}{L} 2S\sum^L_{j=1}j \right)
 \notag\\
 &= \exp(-2\pi i l S (L+1))\hat{U}^{\rm tw}_{2\pi l}
   =(-1)^{2S l}\hat{U}^{\rm tw}_{2\pi l},
\end{align}
where we have used the condition $SL$: integer.
Similarly, we calculate
\begin{align}
 \hat{K}^{\dagger}\hat{U}^{\rm twl}_{2\pi l}\hat{K}
 =\exp(-2\pi i l SL ) \hat{U}^{\rm twl}_{2\pi l}
 = \hat{U}^{\rm twl}_{2\pi l}.
\end{align}

\end{document}